\documentclass[conference]{IEEEtran}

\usepackage[T1]{fontenc}

\usepackage{graphicx}
\usepackage[export]{adjustbox}
\usepackage[dvipsnames]{xcolor}
\usepackage{bbm}
\usepackage{kotex} 

\graphicspath{{./Figs/}}
\usepackage{subfigure}
\usepackage{stfloats}
\ifCLASSINFOpdf
\else
\fi

\usepackage{amssymb}
\usepackage{amsmath}
\usepackage[cmintegrals]{newtxmath}
\usepackage{stackengine}

\usepackage{algorithmic}
\usepackage{algorithm}

\usepackage{amsthm}
\usepackage{xpatch}

\xpatchcmd{\proof}{\hskip\labelsep}{\hskip5\labelsep}{}{}

\usepackage{mathtools}

\usepackage{cite}

\usepackage{array}

\usepackage{booktabs}
\usepackage{multirow}

\begin{document}
\title{UV-Plane Beam Mapping for Non-Terrestrial Networks in 3GPP System-Level Simulations}
\author{\IEEEauthorblockN{Dong-Hyun Jung\IEEEauthorrefmark{1}, Sucheol Kim\IEEEauthorrefmark{1}, Miyeon Lee\IEEEauthorrefmark{2}, Joon-Gyu Ryu\IEEEauthorrefmark{1}, and 
 Junil Choi\IEEEauthorrefmark{2}}
\IEEEauthorblockA{
\IEEEauthorrefmark{1}Satellite Communication Research Division, Electronics and Telecommunications Research Institute, Daejeon, Korea\\
\IEEEauthorrefmark{2}School of Electrical Engineering, Korea Advanced Institute of Science and Technology, Daejeon, Korea\\
E-mail: \IEEEauthorrefmark{1}\{dhjung, loehcusmik, jgryurt\}@etri.re.kr; \IEEEauthorrefmark{2}\{mylee0031, junil\}@kaist.ac.kr
}}
\maketitle

\begin{abstract}
Due to the high altitudes and large beam sizes of satellites, the curvature of the Earth's surface can impact system-level performance. To consider this, 3GPP introduces the UV-plane beam mapping for system-level simulations of non-terrestrial networks (NTNs). This paper aims to provide a comprehensive understanding of how beams and user equipments (UEs) are placed on the UV-plane and subsequently mapped to the Earth's surface. We present a general process of projecting UEs on the UV-plane onto the Earth's surface. This process could offer a useful guideline for beam and UE deployment when evaluating the system-level performance of NTNs.


\end{abstract}

\IEEEpeerreviewmaketitle

\def\tcr{\textcolor{red}}
\def\tcb{\textcolor{blue}}
\def\tcg{\textcolor{green}}

\def\E{\mathbb{E}}
\def\P{\mathbb{P}}

\def\re{r_{\mathrm{E}}}

\def\th{\mathrm{th}}
\def\max{\mathrm{max}}
\def\min{\mathrm{min}}
\def\out{\mathrm{out}}
\def\sec{\mathrm{sec}}
\def\ml{\mathrm{ml}}
\def\sl{\mathrm{sl}}

\def\O{\mathrm{O}}
\def\Ps{\mathrm{P}_{\mathrm{s}}}
\def\Piuv{\mathrm{P}_{\mathrm{u}}^{\mathrm{uv}}}
\def\Pu{\mathrm{P}_{\mathrm{u}}}
\def\duv{d_{\mathrm{u}}^{\mathrm{uv}}}
\def\d{d_{\mathrm{u}}}
\def\thezod{\theta_{\mathrm{LOS,ZOD}}}
\def\theaod{\theta_{\mathrm{LOS,AOD}}}


\section{Introduction}\label{Sec:Intro}
\IEEEPARstart{S}{ystem}-level simulation is an important process for evaluating system performance, such as user equipment (UE) throughput, total cell throughput, and cell-edge throughput. 
A stochastic geometry-based approach was widely used to evaluate the system performance in a very effective way and to check the impacts of various system parameters on the performance \cite{book09SG}. 
Various system-level simulations have been conducted using stochastic geometry in terrestrial networks (TNs) \cite{SG09Haenggi,SG13Singh,SG14Singh,SG17Andrews} and non-terrestrial networks (NTNs) \cite{SG20OkatiBPP,my22TCOM,my23VTM,my23TWCsub,my24TCOMsub}.

Unlike the stochastic model-based system performance evaluation, 3GPP system-level simulations 
assume a deterministic hexagonal grid for cell or beam layout and consider physical, medium access control, and transport layer protocols to estimate more realistic system performance.
To evaluate the system performance of physical layer techniques, 3GPP proposed the cluster-based fast fading channel model in \cite{TR38.901}, which also includes a detailed methodology for calibrating the channel model in specific scenarios.
A guideline to establish satellite channels was provided in \cite{TR38.811} for NTNs. 
In \cite{TR38.821}, the NTN system environments and parameters for link- and system-level performance evaluation were proposed in thirty different study cases assuming single satellite scenarios.



As conventional TNs have considered a flat deployment surface, UEs are distributed within hexagonal beams or cells on the 2D plane. However, in NTNs, due to the high altitudes and large beam sizes of satellites, the curvature of the Earth's surface might affect the system-level performance, so that this should be taken into account in the NTN beam layout. Therefore, in \cite{TR38.821}, the UV-plane is defined in the satellite reference frame, and hexagonal beams are placed on that plane. This paper aims to provide a comprehensive understanding of how beams and UEs are placed on the UV-plane and then mapped to the Earth's surface. And also we hope to offer useful guidelines for beam and UE deployment in NTN system-level simulations.



\begin{figure}
\begin{center}
\includegraphics[width=
.8\columnwidth]{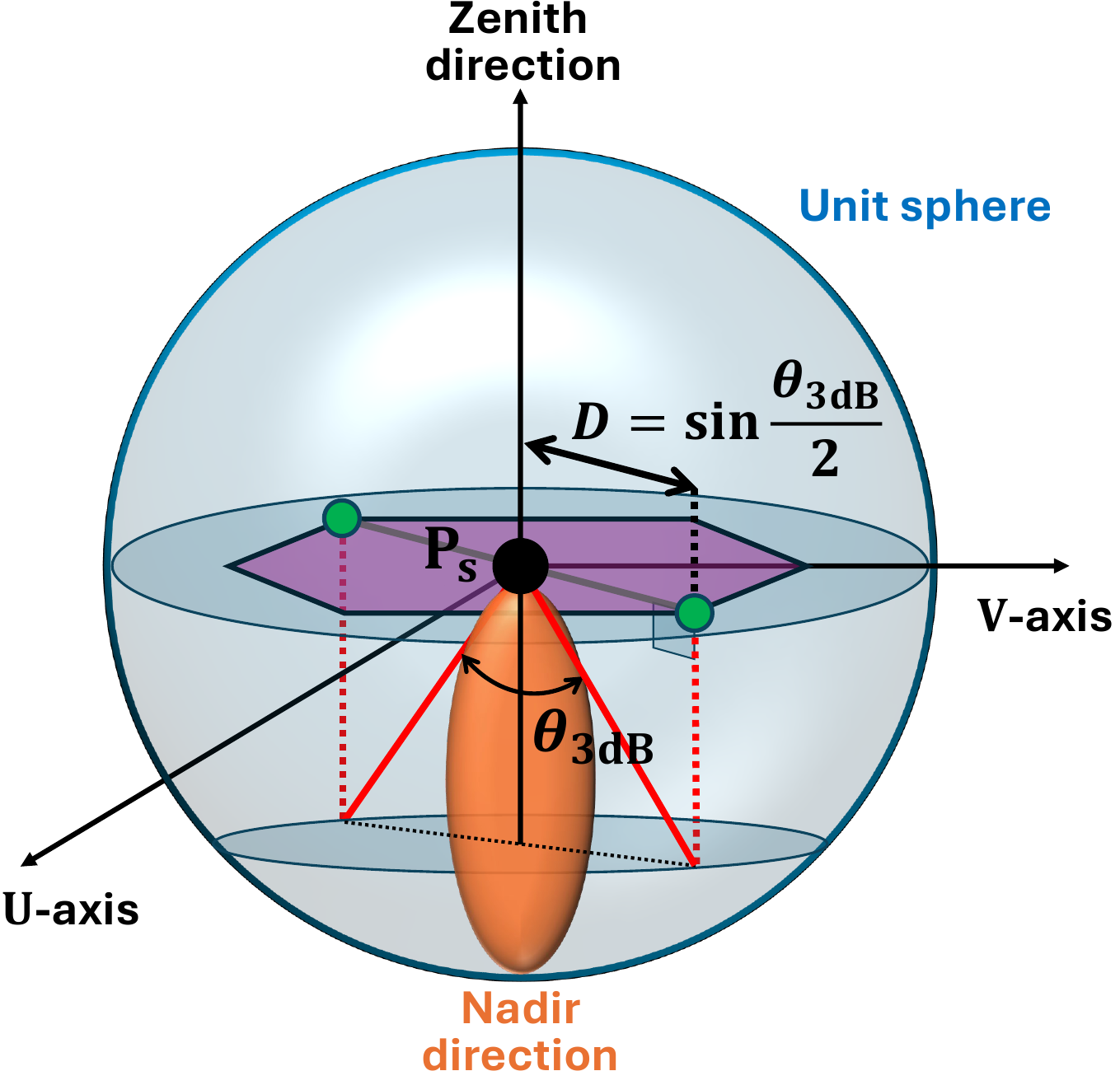}
\end{center}
\setlength\abovecaptionskip{.25ex plus .125ex minus .125ex}
\setlength\belowcaptionskip{.25ex plus .125ex minus .125ex}
\caption{Principle of the UV-plane beam mapping for the 3dB beamwidth $\theta_{\mathrm{3dB}}$ where $\Ps$ is the satellite position, and $D$ is the beam radius in the UV-plane. The central beam center is at nadir point.}
\label{Fig:beam_mapping}
\end{figure}

\section{3GPP NTN Beam Layout}\label{sec:syst_model}
\begin{figure}
\begin{center}
\includegraphics[width=.7\columnwidth]{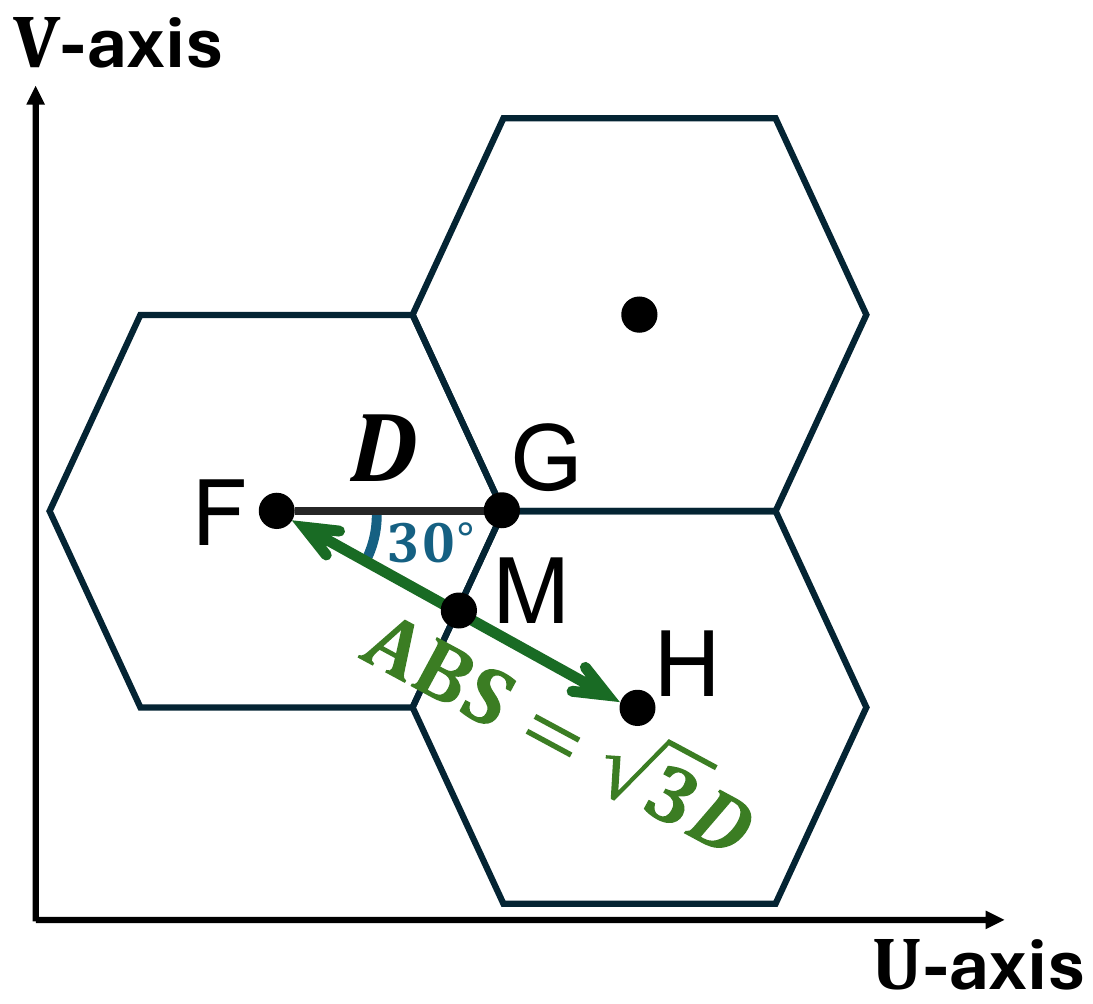}
\end{center}
\setlength\abovecaptionskip{.25ex plus .125ex minus .125ex}
\setlength\belowcaptionskip{.25ex plus .125ex minus .125ex}
\caption{Relationship between the beam radius $D$ and the ABS where $\text{F}$ and $\text{H}$ are beam centers, and $\text{G}$ is the intersection point of the three hexagons' vertices. }
\label{Fig:abs}
\end{figure}

For NTN system-level simulations, 3GPP has established the following beam layout assumptions for a single satellite simulation \cite{TR38.821}.
\begin{itemize}
    \item Hexagonal mapping of the beam boresight directions on the UV-plane is assumed.
    \item The 3dB beamwidth is used to determine the beam diameter and adjacent beam spacing (ABS).
    \item A 19-beam layout that incorporates a wrap-around mechanism: 18 beams around the central beam, surrounded by two distinct interfering tiers.
\end{itemize}
Based on these assumptions, we discuss the principle of the UV-plane beam mapping hereafter. 
As shown in Fig. \ref{Fig:beam_mapping}, the UV-plane is defined as the plane that includes the position of a satellite  and is perpendicular to the Nadir direction, i.e., the direction toward the Earth's center from the satellite, where the U- and V-axes are the orthogonal basis spanning this plane. 
The beam layout projected onto the UV-plane is determined by a unit sphere and the antenna pattern of each beam.
In Fig. \ref{Fig:beam_mapping}, the purple region shows the center beam area spanned in the UV-plane considering the 3dB beamwidth $\theta_{\mathrm{3dB}}$. Thus, the beam radius, denoted by $D$, is readily calculated using the angle $\theta_{\mathrm{3dB}}$ and the characteristics of the unit sphere as
\begin{align}
    D=\sin \frac{\theta_{\mathrm{3dB}}}{2}.
\end{align}
The relationship between the beam radius $D$ and the ABS is shown in Fig. \ref{Fig:abs}. Due to the properties of a regular hexagon, $\angle \text{GFM}$ is equal to 30 degrees. By using $\overline{\text{FG}}=D = \sin \frac{\theta_{\mathrm{3dB}}}{2}$, the ABS, which is the same as $\overline{\text{FH}}$, is given by
\begin{align}\label{eq:abs}
    \text{ABS} = 2 \times \overline{\text{FM}} = 2 \times \frac{\sqrt{3}}{2}D = \sqrt{3} \sin \frac{\theta_{\mathrm{3dB}}}{2}.
\end{align}
From \eqref{eq:abs}, we can calculate ABSs for several scenarios as in Table \ref{table:abs}, which aligns with the results in \cite{TR38.821}. Once the beam layout is configured using the ABS derived in \eqref{eq:abs}, a pre-defined number of UEs are randomly distributed within each hexagonal beam on the UV-plane for system-level performance evaluation. Each UE on the UV-plane should then be mapped to a position on the Earth's surface by utilizing the definition and characteristics of the UV-plane. In the next section, we present a general process of mapping UEs from the UV-plane to positions on the Earth's surface.

\begin{table}[t]
    \centering
    \caption{Beamwidth and ABS for the scenarios in \cite{TR38.821}}
    \label{table:abs}
    \begin{tabular}{llcc}
        \toprule
        \textbf{Set} & \textbf{Scenarios} & \textbf{Beamwidth [deg]} & \textbf{ABS} \\
        \midrule
        \multirow{4}{*}{\textbf{Set-1}} & GEO S-band & 0.4011 & 0.0061 \\
                                        & GEO Ka-band & 0.1765 & 0.0027 \\
                                        & LEO S-band & 4.4127 & 0.0667 \\
                                        & LEO Ka-band & 1.7647 & 0.0267 \\
        \midrule
        \multirow{4}{*}{\textbf{Set-2}} & GEO S-band & 0.7353 & 0.0111 \\
                                        & GEO Ka-band & 0.4412 & 0.0067 \\
                                        & LEO S-band & 8.832 & 0.1334 \\
                                        & LEO Ka-band & 4.4127 & 0.0667 \\
        \bottomrule
    \end{tabular}
\end{table}

\begin{figure}
\begin{center}
\includegraphics[width=\columnwidth]{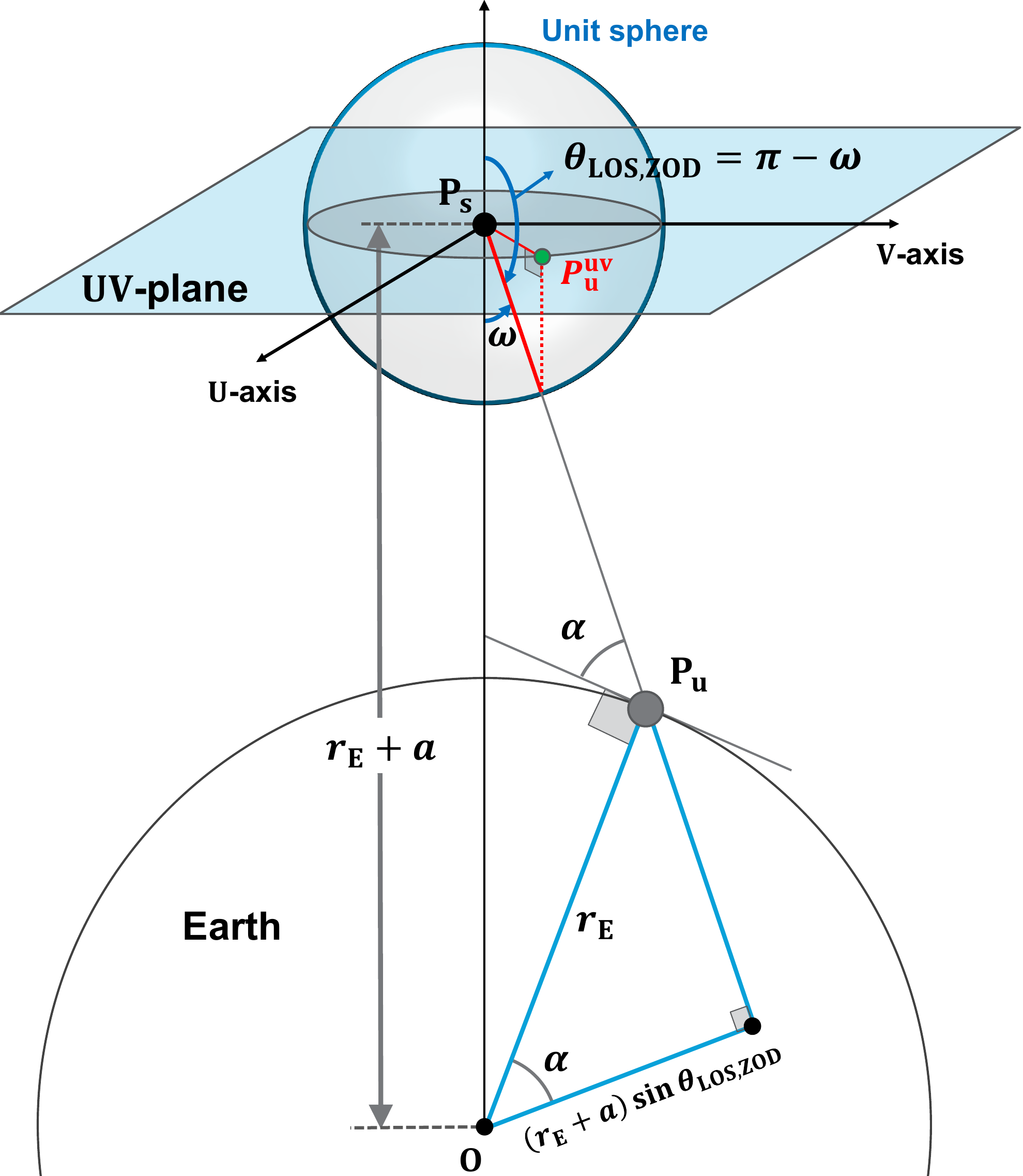}
\end{center}
\setlength\abovecaptionskip{.25ex plus .125ex minus .125ex}
\setlength\belowcaptionskip{.25ex plus .125ex minus .125ex}
\caption{Description of the relationship between the UE position in the UV-plane $\Piuv$ and the actual UE position $\Pu$ where $\mathrm{O}$ is the Earth's center. }
\label{Fig:UE_location_uv_plane}
\end{figure}

\begin{figure*}
\centering
\subfigure[UV-plane, $\text{FRF}=1$]{
\includegraphics[width=\columnwidth]{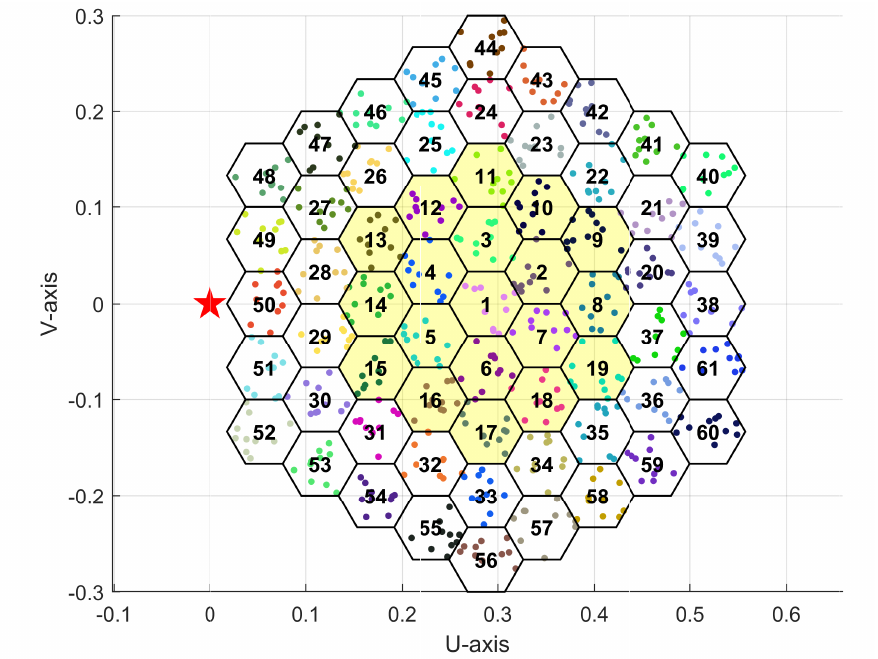}
\label{Fig:beam_layout_uv_FRF1}
}
\subfigure[Earth's surface (top view), $\text{FRF}=1$]{
\includegraphics[width=.9\columnwidth]{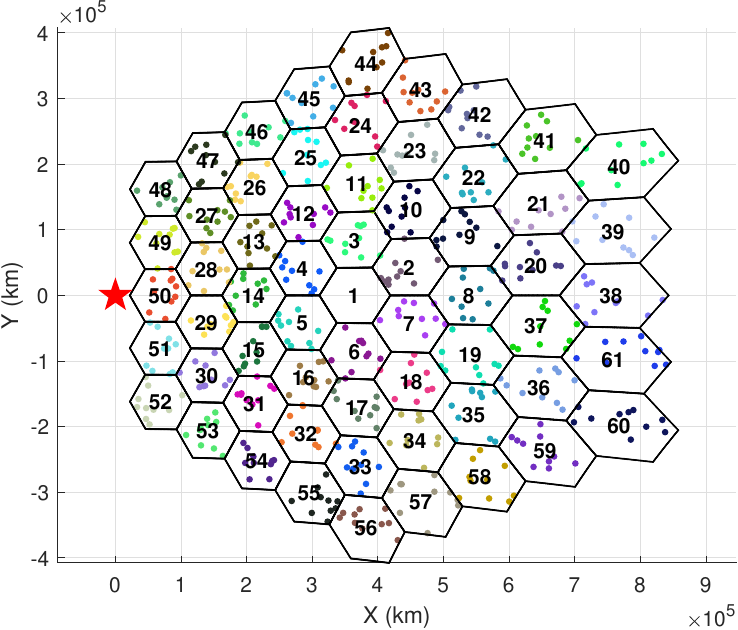}
\label{Fig:beam_layout_Earth_FRF1}
}
\subfigure[UV-plane, $\text{FRF}=3$ ]{
\includegraphics[width=\columnwidth]{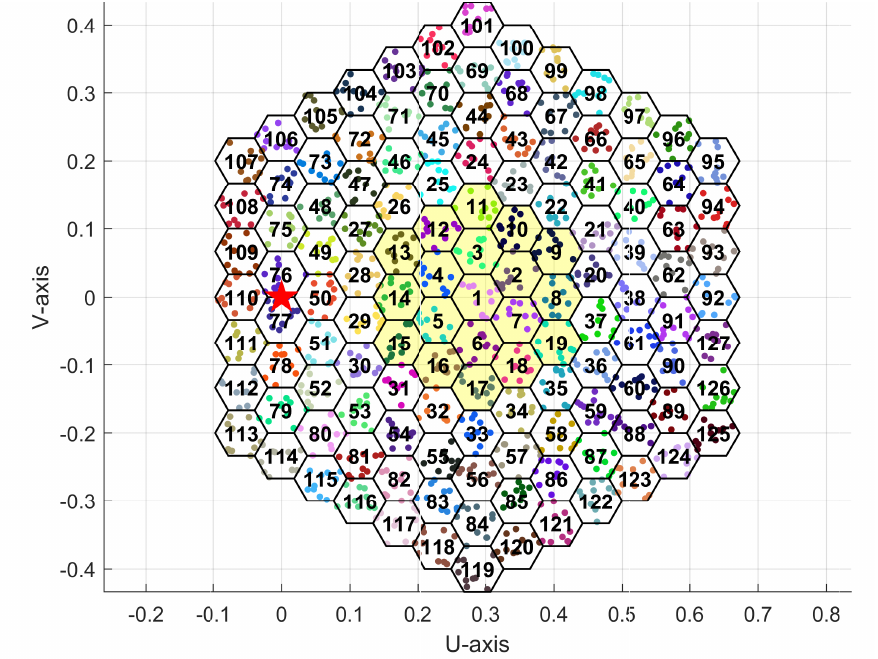}
\label{Fig:beam_layout_uv_FRF3}
}
\subfigure[Earth's surface (top view), $\text{FRF}=3$]{
\includegraphics[width=.9\columnwidth]{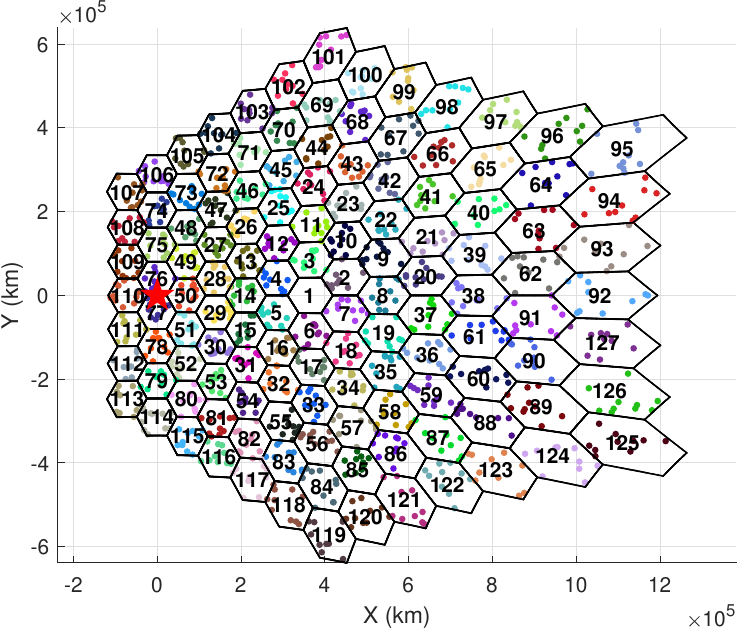}
\label{Fig:beam_layout_Earth_FRF3}
}
\caption{Beam layout (black solid lines) and UE locations (colored dots) on both the UV-plane and the Earth's surface for $\text{FRF}=1$ and $\text{FRF}=3$.
A LEO satellite with Set-1 parameters operating in the S-band is considered where the center beam elevation of 70 degrees. The red stars represent the satellite positions in the UV-plane, and the yellow hexagons are beams for collecting statistics.}\label{Fig:beam_layout_Earth}
\end{figure*}

\section{Projection from UV-plane to Earth's surface}\label{sec:projection}
Let $r_{\mathrm{E}}$ be the Earth's radius and $a$ denote the satellite altitude.
Without loss of generality, we set the satellite position in the Cartesian coordinates as 
\begin{align}
\Ps=[0,0,\re+a]^{\mathrm{T}}    
\end{align}
and focus on one UE whose position on the UV-plane is denoted by $\Piuv$, and that on the Earth surface by $\Pu$.
The important geometry is shown in Fig. \ref{Fig:UE_location_uv_plane}, and the process of placing the UE on the Earth's surface is described as follows.

\begin{enumerate}
    \item Calculate the angle $\omega$ between $\overline{\O \Ps}$ (the line between the Earth's center and the satellite) and $\overline{\Ps \Pu}$ (the line between the satellite and the actual UE position on the Earth's surface). Let $\duv$ denote the distance between the satellite and the UE position in the UV-plane, i.e., 
    \begin{align}
    \duv=\lVert \Ps - \Piuv \rVert.    
    \end{align}
    Then, the angle $\omega$ is obtained by the characteristics of the unit sphere as
    \begin{align}
        \omega=\arcsin \duv.
    \end{align}
    \item Calculate the zenith of departure (ZOD) angle of the line-of-sight (LOS) path using the angle $\omega$ as
    \begin{align}
        \thezod = \pi - \omega.
    \end{align}
    \item Calculate the azimuth of departure (AOD) angle of the LOS path as
    \begin{align}
        \theaod = \angle \Piuv \Ps \hat{\mathbf{u}}
    \end{align}
    where $\hat{\mathbf{u}}$ is the unit vector toward the U-axis.
    \item Using the geometry of the blue triangle in Fig. \ref{Fig:UE_location_uv_plane}, calculate the elevation angle as
    \begin{align}
        \alpha = \arccos\left(\frac{(\re+a)\sin\thezod}{\re}\right).
    \end{align}
    \item Calculate the slant range as
    \begin{align}
        \d = \overline{\Ps \Pu} = -\re\sin\alpha + \sqrt{\re^2\sin^2\alpha + a^2 +2 \re a}.
    \end{align}
    \item Obtain the UE projection vector $\overrightarrow{\Ps \Pu}$ by converting the spherical coordinates with the ZOD and AOD angles into the Cartesian coordinates with a radius of $\d$, i.e.,
    \begin{align}
        \Delta \mathrm{P} = \begin{bmatrix}
                \d \sin (\thezod) \cos (\theaod) \\
                \d \sin (\thezod) \sin (\theaod)\\
                \d \cos (\thezod)
                \end{bmatrix}.    
    \end{align}
    \item Calculate the UE position as $\Pu=\Ps+\Delta \mathrm{P}$.
\end{enumerate}

From the above process, we can project all UEs in the UV-plane onto their actual positions on the Earth's surface. With these actual positions, we can proceed with the system-level simulations, including large- and small-scale channel generation and reference signal received power-based UE attachment.

\begin{figure}
\centering
\subfigure[$\text{FRF}=1$]{
\includegraphics[width=0.95\columnwidth]{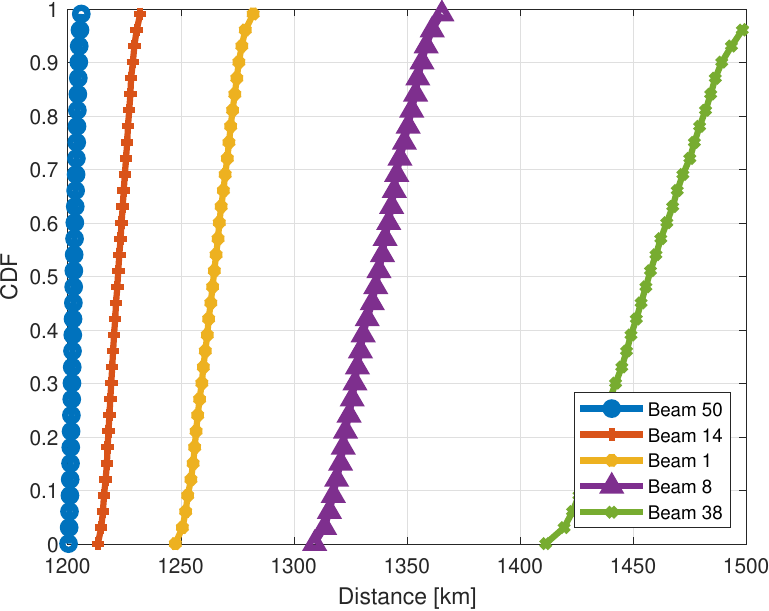}
\label{Fig:dist_dist_FRF1}
}
\subfigure[$\text{FRF}=3$]{
\includegraphics[width=0.95\columnwidth]{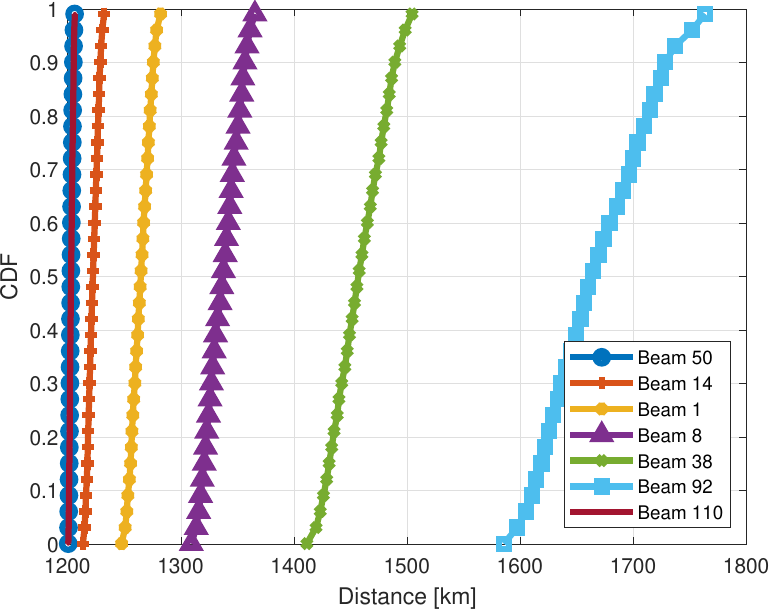}
\label{Fig:dist_dist_FRF3}
}
\caption{Distributions of satellite-UE distances.}
\label{Fig:dist_dist}
\end{figure}
\vspace{-10pt}

\section{Simulation Results}\label{sec:sim_res}






In this section, we simulate the UE projection process presented in the previous section. 
In our scenario, a LEO satellite is located at an altitude of 1200 km with the center beam elevation of 70 degrees. We assume the satellite configuration is based on Set-1 and the S-band, which includes a 3dB beamwidth of 4.4127 degrees and an ABS of 0.0667 in the UV-plane.

Fig. \ref{Fig:beam_layout_Earth} shows the beam layout and UE locations in both the UV-plane and the Earth's surface for the scenario of a LEO satellite with Set-1 parameters operating in the S-band, and the center beam elevation $\theta_{\mathrm{c}}$ of 70 degrees. The numbers of beams for $\text{FRF}=1$ and $\text{FRF}=3$ are 61 and 127, respectively, and ten UEs are randomly dropped within each beam. For system-level simulations, the inner-19 beams (yellow hexagons) are used to collect UE performance statistics, while the outer beams only generate interference to the inner beams. The center of the center beam is located at $(\sin \theta_{\mathrm{c}}, 0)=(0.2878,0)$ in the UV-plane.
As expected, the shape of the beam layout in the UV-plane is distorted when being projected onto the Earth's surface due to the curvature of the Earth’s surface. It can be seen that the actual beam size on the Earth's surface is highly dependent on the number of beams per satellite.
When a satellite generates more beams, the edge beams have higher elevation angle, resulting in a more distorted beam shape.

Fig. \ref{Fig:dist_dist} illustrates the distributions of distances between the satellite and UEs. The UEs in the beams closer to the satellite exhibit the shorter distances due to their higher elevation angles. 
For example, for $\text{FRF}=1$, the UEs in the beam 50 have the shortest distances, while those in the beam 38 have much larger distances. For $\text{FRF}=3$, the edge beams are more distorted due to the longer distance from the center beam. Thus,  distances are spread over a larger range, e.g., from 1200 km (for UEs in the beam 76 or 77) to 1771 km (for UEs in the beam 92) compared to the case of $\text{FRF}=1$.



\section{Conclusions}\label{sec:conclusions}
This paper considered the methodology of UV-plane beam mapping used in 3GPP NTN system-level simulations. First, we described the definition of the UV-plane and the relationship between the 3dB beamwidth and beam size in the plane. Next, we presented a general process for placing beams and UEs on the Earth's surface. This process could provide a valuable guideline for beam and UE deployment when evaluating the system-level performance of NTNs.

\section*{Acknowledgment}
This work was supported by Institute of Information \& communications Technology Planning \& Evaluation (IITP) grant funded by the Korea government (MSIT) (No.2021-0-00847, Development of 3D Spatial Satellite Communications Technology).


\ifCLASSOPTIONcaptionsoff
  \newpage
\fi

\bibliographystyle{IEEEtran}
\bibliography{
    references/3GPP,
    references/books,
    references/chSR,
    references/IEEEabrv, 
    references/myPapers,
    references/refs,
    references/SG
    }

\begin{thebibliography}{10}
\providecommand{\url}[1]{#1}
\csname url@samestyle\endcsname
\providecommand{\newblock}{\relax}
\providecommand{\bibinfo}[2]{#2}
\providecommand{\BIBentrySTDinterwordspacing}{\spaceskip=0pt\relax}
\providecommand{\BIBentryALTinterwordstretchfactor}{4}
\providecommand{\BIBentryALTinterwordspacing}{\spaceskip=\fontdimen2\font plus
\BIBentryALTinterwordstretchfactor\fontdimen3\font minus \fontdimen4\font\relax}
\providecommand{\BIBforeignlanguage}[2]{{%
\expandafter\ifx\csname l@#1\endcsname\relax
\typeout{** WARNING: IEEEtran.bst: No hyphenation pattern has been}%
\typeout{** loaded for the language `#1'. Using the pattern for}%
\typeout{** the default language instead.}%
\else
\language=\csname l@#1\endcsname
\fi
#2}}
\providecommand{\BIBdecl}{\relax}
\BIBdecl

\bibitem{book09SG}
F.~Baccelli and B.~Blaszczyszyn, \emph{Stochastic geometry and wireless networks: Volume I theory}.\hskip 1em plus 0.5em minus 0.4em\relax Found. Trends in Networking, 2009.

\bibitem{SG09Haenggi}
M.~Haenggi, J.~G. Andrews, F.~Baccelli, O.~Dousse, and M.~Franceschetti, ``Stochastic geometry and random graphs for the analysis and design of wireless networks,'' \emph{{IEEE} J. Sel. Areas Commun.}, vol.~27, no.~7, pp. 1029--1046, 2009.

\bibitem{SG13Singh}
S.~Singh, H.~S. Dhillon, and J.~G. Andrews, ``Offloading in heterogeneous networks: Modeling, analysis, and design insights,'' \emph{{IEEE} Trans. Wireless Commun.}, vol.~12, no.~5, pp. 2484--2497, 2013.

\bibitem{SG14Singh}
S.~Singh and J.~G. Andrews, ``Joint resource partitioning and offloading in heterogeneous cellular networks,'' \emph{{IEEE} Trans. Wireless Commun.}, vol.~13, no.~2, pp. 888--901, 2014.

\bibitem{SG17Andrews}
J.~G. Andrews, T.~Bai, M.~N. Kulkarni, A.~Alkhateeb, A.~K. Gupta, and R.~W. Heath, ``Modeling and analyzing millimeter wave cellular systems,'' \emph{{IEEE} Trans. Commun.}, vol.~65, no.~1, pp. 403--430, 2017.

\bibitem{SG20OkatiBPP}
N.~Okati, T.~Riihonen, D.~Korpi, I.~Angervuori, and R.~Wichman, ``Downlink coverage and rate analysis of low {Earth} orbit satellite constellations using stochastic geometry,'' \emph{{IEEE} Trans. Commun.}, vol.~68, no.~8, pp. 5120--5134, 2020.

\bibitem{my22TCOM}
D.-H. Jung, J.-G. Ryu, W.-J. Byun, and J.~Choi, ``Performance analysis of satellite communication system under the shadowed-rician fading: A stochastic geometry approach,'' \emph{{IEEE} Trans. Commun.}, vol.~70, no.~4, pp. 2707--2721, Apr. 2022.

\bibitem{my23VTM}
D.-H. Jung, G.~Im, J.-G. Ryu, S.~Park, H.~Yu, and J.~Choi, ``Satellite clustering for non-terrestrial networks: Concept, architectures, and applications,'' \emph{{IEEE} Veh. Technol. Mag.}, vol.~18, no.~3, pp. 29--37, Sep. 2023.

\bibitem{my23TWCsub}
D.-H. Jung, H.~Nam, J.~Choi, and D.~J. Love, ``Modeling and analysis of {GEO} satellite networks,'' 2023, arXiv:2312.15924.

\bibitem{my24TCOMsub}
M.~Lee, S.~Kim, M.~Kim, D.-H. Jung, and J.~Choi, ``Analyzing downlink coverage in clustered low {Earth} orbit satellite constellations: A stochastic geometry approach,'' 2024, arXiv:2402.16307.

\bibitem{TR38.901}
{3GPP TR 38.901 v14.3.0}, ``Study on channel model for frequencies from 0.5 to 100 {GHz},'' Jan. 2018.

\bibitem{TR38.811}
{3GPP TR 38.811 v15.4.0}, ``Study on {NR} to support non-terrestrial networks,'' Sep. 2020.

\bibitem{TR38.821}
{3GPP TR 38.821 v16.0.0}, ``Solutions for {NR} to support non-terrestrial networks ({NTN}),'' Dec. 2019.

\end{thebibliography}

\end{document}